\documentstyle[aps,preprint]{revtex}
\begin{document}
 \draft
\title{Nuclear symmetry energy in presence of hyperons in the nonrelativistic 
Thomas-Fermi approximation}
\author{Sarmistha Banik and Debades Bandyopadhyay } 
\address{Saha Institute of Nuclear Physics, 1/AF Bidhannagar, 
Calcutta 700 064, India}

\maketitle

\begin{abstract}
We generalise the finite range momentum and density dependent Seyler-Blanchard
nucleon-nucleon effective interaction to the case of interaction between two 
baryons. This
effective interaction is then used to describe dense hadronic matter relevant
to neutron stars in the nonrelativistic Thomas-Fermi approach. We investigate
the behaviour of nuclear symmetry energy in dense nuclear and hyperon matter
relevant to neutron stars. It is found that the nuclear symmetry energy always
increases with density in hyperon matter unlike the situation in nuclear
matter. This rising characteristic of the symmetry energy in presence of
hyperons may have significant implications on the mass-radius relationship
and the cooling properties of neutron stars. We have also noted that with the 
appearance of hyperons, the equation of state calculated in this model remains
causal at high density.
\end{abstract}

\newpage

The study of matter far off from normal nuclear matter density is of interest 
in understanding various of properties of neutron stars. The matter density
in the core of neutron stars could exceed up to a few times normal nuclear
matter density. Our knowledge about dense matter 
is very much constrained by a single density point 
in the whole density plane i.e. normal nuclear matter density or the saturation
density. The empirical values of various properties of symmetric nuclear
matter i.e. binding energy, bulk symmetry energy, compressibility are only 
known at this density. All models are fitted to those properties at the 
saturation density and then extrapolated to high density regime.

The symmetry energy is an essential input in understanding gross properties
of neutron stars. The bulk symmetry energy is
defined as the difference between the energy per particle for pure neutron
matter and that of symmetric nuclear matter at normal nuclear matter density.
The empirical value of the bulk symmetry energy lies in the range 30-40 MeV.
Nuclear symmetry energy controls Fermi momenta of baryons, particle fractions
and the equation of state of dense matter. Since a dense system like neutron
stars is an infinite one, the volume and symmetry energy terms in the
Bethe-Weizs\"acker mass formula contribute to the total energy of the system.
As a consequence, the energy of the system is lowered when the system is more
symmetric i.e. its symmetry energy is less.

Though various nonrelativistic as well as relativistic models are fitted to 
the symmetry energy at the saturation 
density, there is no consensus among the models about the behaviour of nuclear
symmetry energy far off from normal nuclear matter density. It was earlier
noted by many authors \cite{Wir,Kut1,Kut2} that the symmetry energy, in 
nonrelativistic models,
initially increased and afterwards it either decreased with density or 
saturated to a value. It was attributed to the role of the tensor interaction
\cite{Wir,Kut1,Kut2,Pan} in isospin
singlet (T=0) nucleon pairs. At low density, the attraction due to the tensor
force dominates over the short range repulsion in T=0 nucleon pairs. As a
consequence, symmetric nuclear matter is more attractive than pure neutron
matter and the symmetry energy increases with density initially. At high 
density, the tensor interaction in T=0 channel vanishes \cite{Pan} and the 
short range
repulsion in isospin singlet nuclear pairs wins over that of isospin triplet
pairs. As a result, nuclear symmetry energy falls at high density regime
leading to  energetically favourable pure neutron matter and the 
disappearance of protons. It was shown by Engvik et al. \cite{Eng}
that the symmetry energy increased with density in the lowest order Brueckner
calculations using modern nucleon-nucleon (NN) potentials. Such a behaviour
of the symmetry energy was also reported in another Brueckner calculation
using realistic nucleon-nucleon potentials \cite{Bal}. On the other hand,
Akmal et al. \cite{Akm} found that the symmetry increased at lower densities
and then decreased at high
density in variational chain summation (VCS) method using one such modern 
NN potential i.e. A18. The difference between those calculations may be
stemmed from the neglect of higher order terms in Brueckner calculations.
Akmal et al. also observed that the proton fraction calculated in the VCS 
approach using A18 plus three nucleon interaction, increased with density.
However, they noted that the too strong repulsion in the three nucleon force
resulted in overestimation in the proton fraction or the symmetry energy.

In relativistic mean field (RMF) models \cite{Kut2,Gle}, the symmetry energy 
always increases with density. Here, the mean $\rho$-meson field is responsible 
for the interaction part of the symmetry energy and it increases with density. 
However, two main features $-$ the tensor force and different repulsive 
strengths in isospin singlet and isospin triplet nucleon pairs, are absent 
in RMF calculations. 

In various nonrelativistic models, the fall of the symmetry energy occurs 
beyond a few times normal nuclear matter density. On the other hand, the formation of hyperons
is a possibility at about 2-3 times normal nuclear matter density. Therefore,
it may be a serious flaw to consider a dense matter system consisting only of 
nucleons at high density. Also, nonrelativistic models consisting only of 
nucleons violate causality at high density. This problem might be rectified 
with the appearance of hyperons. Hyperons are created at the cost of nucleons'
energy. With the formation of hyperons, Fermi momenta (velocities) of nucleons
will be reduced. On the other hand, hyperons being heavier than nucleons will
have smaller Fermi velocities. In this situation, all baryons may
be treated as nonrelativistic particles in a dense system. Strange hadron 
systems were studied extensively in RMF models \cite{Gle,Sch1,Sch2,Gal}. 
Recently, there have been some calculations on strange hadronic matter
in the nonrelativistic Brueckner approximation using baryon-baryon 
potentials \cite{Bal,Sto} and also using phenomenologically constructed energy
density functional \cite{Balb}. 

In this letter, we investigate the density dependence of nuclear symmetry 
energy in the
nonrelativistic Thomas-Fermi approximation using a momentum and density
dependent finite range Seyler-Blanchard effective interaction. 
The momentum dependent Seyler-Blanchard nucleon-nucleon effective interaction 
was extensively 
applied to the determination of the parameters of mass formula by Myers and
Swiatecki \cite{Mye1}. However, the energy dependence of the single particle 
potential was too strong because of the strong momentum dependence
in the effective interaction. Later, the momentum dependent 
Seyler-Blanchard effective interaction was modified to include 
a two-body density dependent term which simulated three body effects and the
energy dependence of the single particle potential was exploited to delineate
the momentum and density dependence of the effective interaction 
\cite{Ban1,Ban2,Mye2}. 
This modified Seyler-Blanchard (SBM) interaction was used in the description 
of heavy ion collisions \cite{Ban1}, dense matter properties \cite{Ban2,Mye2}
and neutron stars \cite{Mah}. Here, we generalise the SBM interaction to the 
case of baryon-baryon interaction with the inclusion of hyperons in addition to
nucleons. Later, we exploit this momentum and density dependent finite range
baryon-baryon effective interaction to calculate nuclear symmetry energy in
nuclear and hyperon matter relevant to neutron stars.

The interaction between two baryons with separation $r$ and relative 
momentum $p$ is given by
\begin{equation}
V_{eff}(r,\rho,p)= -C_{B_{1}B_{2}}[1-\frac{p^2}{b^2}-d^2({\rho_1 + \rho_2})^n]
{\frac {e^{-r/a}}{r/a}},
\end{equation}
where $a$ is the range parameter and  
$b$ defines  the strength of repulsion in the momentum space; $d$ 
and $n$ are two parameters determining the strength of the density dependence; 
$\rho_1$ and $\rho_2$ are total baryon densities at the sites of two 
interacting baryons. We have all the baryons of SU(3) octet and leptons($e^{-}$,
$\mu^{-}$) in our calculation. The constituents of matter in neutron stars
are highly degenerate and the chemical potentials of baryons and leptons are
much larger than the temperature of the system. Therefore, our calculation is 
confined to zero temperature case. The single particle potential for baryon 
$B_{1}$ is defined as,
\begin{eqnarray}
V_{B_{1}}(p_{1},\rho)&=&V_{B_{1}}^{0}+p_{1}^2 V_{B_{1}}^{1}+V^2 \nonumber\\
&=&{2 \over (2\pi)^3} \int d\vec{p_{2}}d\vec{r_{2}}V_{eff}[C_{B_{1}B_{1}}
\Theta(p_{F_{B_1}}
- p_{2})+\sum_{B_{2} \neq B_{1}}C_{B_{1}B_{2}}\Theta(p_{F_{B_2}}- p_{2})+V^2],
\end{eqnarray}
where, $V^0_{B_1}$, $V^1_{B_1}$ are the momentum independent and 
dependent parts of the single particle potential, respectively and 
$V^2$, the rearrangement contribution \cite{Ban2} arising out of 
the density dependence of the two-body effective interaction, is given by
\begin{eqnarray}
V^{2}={1 \over 2}\int d\vec{{r}^{'}}{{\partial v_{2}} \over{\partial{\rho}}}
\sum_{B_{1}} \rho_{B_{1}}[C_{B_{1}B_{1}}\rho_{B_{1}}+\sum_{B_{2} \neq B_{1}}
C_{B_{1}B_{2}}\rho_{B_{2}}],
\end{eqnarray}
with
$v_{2}={d^2(2\rho)^n}{\frac {e^{-r/a}}{r/a}}$. Here, the total baryon density
is denoted by $\rho$ and the summations over $B_1$
and $B_2$ go over all the species of SU(3) baryon octet. 
The density for baryon $B$ is denoted by $\rho_{B}$ and Fermi momentum by 
$ P_{F_{B}}$. 
The effective mass is defined as, 
\begin{equation}
m_{B}^*=[{1 \over m_{B}}+2 V_{B}^1]^{-1}.
\end{equation}
We have from equations (1), (2) and (3)
\begin{eqnarray}
V_{B_{1}}^{0} &=& 4\pi a^{3}(d^{2}(2\rho)^{n}-1)[C_{B_{1}B_{1}}\rho_{B_{1}}+
\sum_{B_{2} \neq B_{1}}C_{B_{1}B_{2}} \rho_{B_{2}}]\nonumber\\
&+& {4a^{3} \over {\pi b^{2}}}[C_{B_{1}B_{1}}{p_{F_{B_1}}^{5} \over 5}
+ \sum_{B_{2} \neq B_{1}}C_{B_{1}B_{2}}{p_{F_{B_2}}^{5} \over 5}],
\end{eqnarray}
\begin{equation}
V_{B_{1}}^{1}={4\pi a^{3}\over {b^2}}[C_{B_{1}B_{1}}\rho_{B_{1}}+\sum_{B_{2} 
\neq B_{1}}C_{B_{1}B_{2}} \rho_{B_{2}}],
\end{equation}
\begin{equation}
V^{2}=4\pi a^{3}d^2 n (2\rho)^{n-1}\sum_{B_{1}}\rho_{B_{1}}[C_{B_{1}B_{1}}
\rho_{B_{1}}+\sum_{B_{2} \neq B_{1}}C_{B_{1}B_{2}}
\rho_{B_{2}}].
\end{equation}
The chemical potential of baryon $B$ is given by,
\begin{eqnarray}
\mu_{B}={P_{F_{B}}^2 \over {2m_{B}^*}} + V_{B}^0 + V^2.
\end{eqnarray}

The symmetry energy is an essential ingredient in understanding dense 
matter. As our calculations are concerned with a system having density far off 
from normal nuclear matter density, it is very much necessary 
to know the behaviour of nuclear symmetry energy at high density.
The energy per nucleon in asymmetric matter may be written as \cite{Lej,Pra1}, 
\begin{equation}
E(\rho,\beta)= E(\rho,{\beta=0} )+ {\beta^2} E_{sym}(\rho), 
\end{equation}
where $\beta = {\frac {(\rho_n - \rho_p)}{\rho}}$ is the asymmetry parameter
and $\rho_n$ and $\rho_p$ are neutron and proton densities, respectively; 
$E(\rho,\beta=0)$ and $E_{sym}(\rho)$ are energy per nucleon in symmetric matter
and nuclear symmetry energy, respectively. It can be shown that the symmetry
energy is related to neutron and proton chemical potentials \cite{Pra1}.
Neutron and proton chemical potentials are defined respectively as  
$\mu_{n}={\partial{\epsilon} \over \partial{\rho_{n}}}$ and 
$\mu_{p}={\partial{\epsilon} \over \partial{\rho_{p}}}$, where 
$\epsilon = {\rho} E(\rho,\beta)$ is the energy density. The expression of 
nuclear symmetry energy follows from equation (9) and the definitions of 
the chemical potentials as,
\begin{eqnarray}
\mu_{n}-\mu_{p}= 4 {\beta} E_{sym}(\rho).
\end{eqnarray}
Putting the expression for chemical potentials (equation (8)) along with 
equations (5)-(7) in equation (10), we obtain 
\begin{equation}
4 \beta E_{sym}(\rho) = E_{kin} + V_{s},
\end {equation}
where the kinetic and the interaction parts of the symmetry energy are 
respectively given by, 
\begin{equation}
E_{kin}= \left[{P_{F_n}^2 \over {2 m_{n}^*}}-{P_{F_p}^2 
\over {2 m_{p}^*}}\right],
\end {equation}
and 
\begin{equation}
V_{s}=[4 \pi a^3 (d^2(2 \rho)^n-1) 
(\rho_{n}-\rho_{p})+{4 a^3 \over {5 \pi b^2}}(p_{F_{n}}^5-p_{F_{p}}^5)]
(C_{nn}-C_{np}).
\end {equation}
\vspace{0.2cm}

The five parameters of nucleon-nucleon interaction in equation (1) - two 
strength parameters $C_{BB}$s (one for $pp$ or $nn$ interaction and the other 
for $np$ or $pn$ interaction), $a$, $b$ and $d$ 
are determined for a fixed value of $n={1/3}$ by reproducing the 
saturation density of normal nuclear matter ($\rho_{0}=0.1533$ $fm^{-3}$), the 
volume energy coefficient for symmetric nuclear matter ($-16.1$ MeV), asymmetry 
energy coefficient ($34$ MeV), the surface energy coefficient of symmetric 
nuclear matter ($18.01$ MeV) and the energy dependence of the real part of the 
nucleon-nucleus optical potential. With the above choice of $n$, the 
incompressibility of normal nuclear matter turns out to be $260$ MeV. 
Also, the effective mass ratio ($m_N^*/m_N$) comes out to be 0.61 at normal
nuclear matter density in our calculation. The values of parameters for 
nucleon-nucleon interaction are presented in Table I. 

Informations about nucleon-hyperon interactions are confined to hypernuclei 
data \cite{Chr}. There is a large body of data on binding energies of 
$\Lambda$-hypernuclei. 
Analyses of those experimental data on hypernuclei indicate that 
the potential felt by a $\Lambda$ in normal nuclear matter is $\simeq -30$ MeV. 
With our two-body baryon-baryon interaction (equation (1)), we determine the 
strength of nucleon-hyperon and hyperon-hyperon interaction from equation (2) 
keeping two range parameters ($a$ and $b$) and the density dependence of 
the interaction same as that of the nucleon-nucleon interaction. 
Parameters of nucleon-$\Lambda$ interaction are shown in Table II. 

Experimental data of $\Sigma$-hypernuclei are scarce and ambiguous because 
of the strong 
$\Sigma$N $\rightarrow \Lambda$N decay. It is also assumed that the 
$\Sigma$ well depth \cite{Sch2} 
in normal nuclear matter is equal to that of a $\Lambda$ particle. 
Therefore, the strength of $\Sigma$N interaction is the same as that of 
$\Lambda$N interaction in our calculation and this is shown in Table II.

In emulsion experiments with $K^{-}$ beams, there are a few events attributed 
to the formation of $\Xi$-hypernuclei. These data can be explained in terms of 
a potential well of $\simeq -25$ MeV for $\Xi$ particle in symmetric nuclear 
matter \cite{Dov}.
We obtain the strength parameter ($C_{BB}$) of $\Xi$N interaction by fitting 
the single particle potential to the above mentioned value and present in 
Table II.

There are a few events of $\Lambda\Lambda$ hypernuclei. Analyses of those 
events indicate a rather strong hyperon-hyperon interaction. 
Schaffner et al. \cite{Sch2} constructed single particle potentials 
on the basis of one boson exchange calculations of Nijmegen group and 
the well depth of a hyperon in hyperon matter is estimated to be 
$\simeq -40$ MeV and this is universal for all hyperon-hyperon 
interactions. The parameters of hyperon-hyperon interaction 
are given in Table II.

In all cases, we notice that interactions involving hyperons, are weaker 
compared to nucleon-nucleon interaction.

The composition of a neutron star is constrained by charge neutrality and 
baryon number conservation. Also, constituents of matter are in 
beta-equilibrium. Baryon chemical potentials are related to neutron and lepton
chemical potentials through the general relation given by
\begin{equation}
\mu_{i}=b_{i}\mu_{n}-q_{i}\mu_{l},
\end{equation}
where $b_{i}$ and $q_{i}$ are the baryon number and charge of i-th baryon 
species, respectively and '$l$' stands for electrons and muons. Solving the 
above mentioned constraints, at a given density, self-consistently, we 
obtain effective masses, Fermi momenta or chemical 
potentials which determine the gross properties of neutron stars.

Particle abundances of nucleons-only matter relevant to  a neutron star are
shown in
figure 1. Here, we notice that the proton (electron) fraction initially 
increases with density and decreases at higher densities. Such a behaviour of 
the proton fraction with density in nonrelativistic models 
was noted earlier by various authors 
\cite{Wir,Kut1,Kut2}. They attributed it to the density dependence of 
nuclear symmetry energy. We will discuss about this later.

In figure 2, particle fractions in hyperon matter relevant to a neutron star 
are plotted with baryon density. The threshold condition for the appearance 
of hyperons depends not only on their masses but also on their charges and 
interaction strengths. The threshold condition is given by
\begin{equation}
\mu_{n}-q_{B}\mu_{e} \geq m_{B}^*+V_{B}^0+V^2,
\end{equation}
where $\mu_{n}$ and $\mu_{e}$ are neutron and electron chemical potentials
respectively, $q_{B}$ is the charge of baryon B. The quantities on the right 
hand side of equation (15) are given by equations (4) - (7). When the 
left hand side equals to or exceeds the right hand side of equation (15), 
baryon species B will be populated. Here, we notice that 
hyperons first appear at 1.5 times normal matter density. Also, it 
is noted that the electron fraction decreases monotonically because 
negatively charged hyperons make the neutron star almost charge neutral. On 
the other hand, the proton fraction is enhanced in hyperon matter compared to 
the situation in nuclear matter ( see figures 1 and 2). 
Moreover, the proton fraction does not show any declining tendency 
at high density as it has been observed in nucleons-only system.
We plot absolute proton density with baryon density in figure 3. The dashed
line (curve $a$) denotes neutron-proton system 
whereas the solid line (curve $b$) represents hyperon system. We 
find that the proton density always increases with baryon density in hyperon 
environment. It may be attributed to the behaviour of 
nuclear symmetry energy with density in hyperon matter.

Violation of causality at high density is a problem in nonrelativistic
models \cite{Akm,Ban2}. The speed of sound ($v^2 = 
{{\partial P} \over {\partial \epsilon}}$) in nucleons-only matter becomes
superluminal i.e. greater than the velocity of light at high density. With the
appearance of additional degrees of freedom in the form of hyperons, Fermi
momenta of neutrons and protons are reduced at high density in comparison to
the situation with nucleons-only matter. As a consequence, the equation of state
now respects causality at densities which might occur at the centers of neutron
stars.

Nuclear symmetry energy is plotted with baryon density in figure 4. 
The dashed line (curve $a$) represents the calculation for nuclear matter 
whereas the solid line (curve $b$) implies that of hyperon matter. We find that 
the nuclear symmetry energy in nuclear matter increases initially with density 
and decreases later at high density. It was pointed out by many authors 
that the fall of the symmetry energy was due to the greater short-range 
repulsion in isospin singlet nucleon pairs than that of 
isospin triplet pairs at high density \cite{Wir,Kut1,Kut2}.
In our calculation , there are two strength parameters in the SBM 
nucleon-nucleon effective interaction i.e. $C_{nn}$ ($C_{pp}$) 
which represents isospin triplet state (T=1) and $C_{np}$ ($C_{pn}$) 
implying isospin singlet (T=0) state. It is evident from Table I that the 
strength parameter ($C_{np}$) in isospin singlet state is stronger 
than that of the triplet state ($C_{nn}$). It is the interaction term
($V_s$) in the symmetry energy (see equation (13)) 
that regulates the behaviour of nuclear symmetry energy. At lower densities, 
the interaction term in $E_{sym}$ is positive because the 
repulsive first term in $V_{s}$ is larger than the attractive second term. 
Therefore, the symmetry energy is increasing 
at lower densities. On the other hand, the interaction term, $V_s$, becomes 
negative around $\sim 4\rho_{0}$  because the second term, in equation (13), 
which is attractive in nature, wins over the first term. Thus at high density, 
pure neutron matter is energetically favourable and protons 
disappear from the system.

We observe that the symmetry energy in nuclear matter ( curve a in figure 4)
starts falling around density $\sim 4\rho_0$. On the other hand, the 
appearance of hyperons is a possibility at about 2-3 times normal nuclear
matter density. Therefore, it may not be justified to consider a system
consisting only of nucleons at high density. Here, we discuss the density
dependence of nuclear symmetry energy including hyperons in our 
nonrelativistic calculation. Nuclear symmetry energy in presence of hyperons 
is calculated using equations (11), (12) and (13).
In hyperon matter, neutrons and protons couple to a hyperon with the same 
coupling strength. Therefore, those terms originating from nucleon-hyperon
interaction cancel out in the calculation of nuclear symmetry energy in 
hyperon matter. The solid line (curve $b$) in figure 4 represents our
calculation of the symmetry energy in hyperon matter. It increases 
with density. This may be attributed to the behaviour of the
interaction term($V_s$) in $E_{sym}$ (see equation (13)) in a hyperon 
environment. Hyperons are produced at the cost of the energy of nucleons. 
Therefore, Fermi momenta of neutrons and protons are 
reduced with the appearance of hyperons compared to the case of nuclear matter. 
As a result, the repulsive first term of $V_s$ (equation (13)) dominates
over the attractive second term leading to a rising nuclear symmetry energy 
for all densities. This behaviour of the symmetry energy in hyperon matter 
is also reflected in the proton fraction (curve b in figure 3). The proton
fraction in neutron stars is crucial in the determining the direct URCA process
which leads to the cooling of neutron stars \cite{Bog,Lat}. The direct URCA
process happens if the proton fraction exceeds the threshold value i.e. 
11 percent. This happens in
our calculation including hyperons around $\sim 3.0 \rho_0$. 

We have compared our nonrelativistic calculations with those of relativistic
mean field models. Nuclear symmetry energy calculated in 
RMF models increases monotonically with density \cite{Kut2}. 
In RMF models, the interaction part of the symmetry energy is related 
to the mean $\rho$ meson field which always
increases with density. It is to be noted that the symmetry energy 
calculated in RMF models rises faster compared to nonrelativistic
calculations \cite{Kut2}. This may be attributed to the fact that the different 
repulsive strengths in isospin singlet and isospin triplet nucleon pairs are
not taken into account by RMF models.
 
In conclusion, we have studied nuclear symmetry energy in nuclear and 
hyperon matter relevant to neutron stars in the nonrelativistic Thomas-Fermi 
approximation using a momentum and density dependent finite range 
Seyler-Blanchard baryon-baryon effective interaction. In nuclear matter, 
the symmetry energy (proton fraction) initially increases
and later it falls with density. With the appearance of hyperons, 
nuclear symmetry energy increases with density in hyperon matter. 
The proton fraction follows the same trend as that of the symmetry energy. 
The increasing symmetry energy 
or proton fraction might have important bearings on the mass-radius 
relationship and cooling properties of neutron stars. 
We will report on these aspects in a future publication.

\newpage
 
\begin{table}
\caption{The different parameters of the effective nucleon-nucleon
interaction as given in equation (1). The parameters -  $a$ and $b$
define the ranges of the interaction in coordinate and momentum space, 
respectively; density dependence of the interaction is given by $d$ and $n$. 
The strengths of $nn$($pp$) and $np$($pn$) interactions are denoted 
by $C_{nn(pp)}$ and $C_{np(pn)}$, respectively.}
\vspace{0.2cm}
\begin{center}
\begin{tabular}{|c|c|c|c|c|c|}\hline
&&&&&\\ 
{n}&{a }&{b}&{d}&{$C_{nn(pp)}$}&{$C_{np(pn)}$}\\ 
\hfil &($fm$)&(MeV)&($fm^{1/2}$)&(MeV)&(MeV)\\
\hline
&&&&&\\
{${1 \over 3}$}&{0.572}&{759.5}&{0.827}&{254.2}&{787.2} \\ \hline
\end{tabular}
\end{center}
\end{table}
\begin{table}
\caption{The different parameters of the effective nucleon-hyperon and
hyperon-hyperon interactions as given in equation (1). The parameters -  $a$ 
and $b$ define the ranges of the interactions in coordinate and momentum space, 
respectively; density dependence of the interactions is given by $d$ and $n$. 
The strengths of $N\Lambda$, $N\Sigma$, $N\Xi$ and $hyperon-hyperon$ 
interactions are denoted 
by $C_{N\Lambda}$, $C_{N\Sigma}$, $C_{N\Xi}$ and $C_{hh}$, respectively.}
\vspace{0.2cm}
\begin{center}
\begin{tabular}{|c|c|c|c|c|c|c|c|}\hline
&&&&&&&\\ 
{n}&{a}&{b}&{d}&{$C_{N \Lambda}$}&{$C_{N \Sigma}$}&{$C_{N \Xi}$}&{$C_{hh}$}\\ 
\hfil &($fm$)&(MeV)&($fm^{1/2}$)&(MeV)&(MeV)&(MeV)&(MeV)\\
\hline
&&&&&&&\\
{${1 \over 3}$}&{0.572}&{759.5}&{0.827}&{262.8}&{262.8}&{233.2}&{462.5}\\ \hline
\end{tabular}
\end{center}
\end{table}
\newpage
{\large{\bf Figure Captions}}
\vspace{1.2cm}

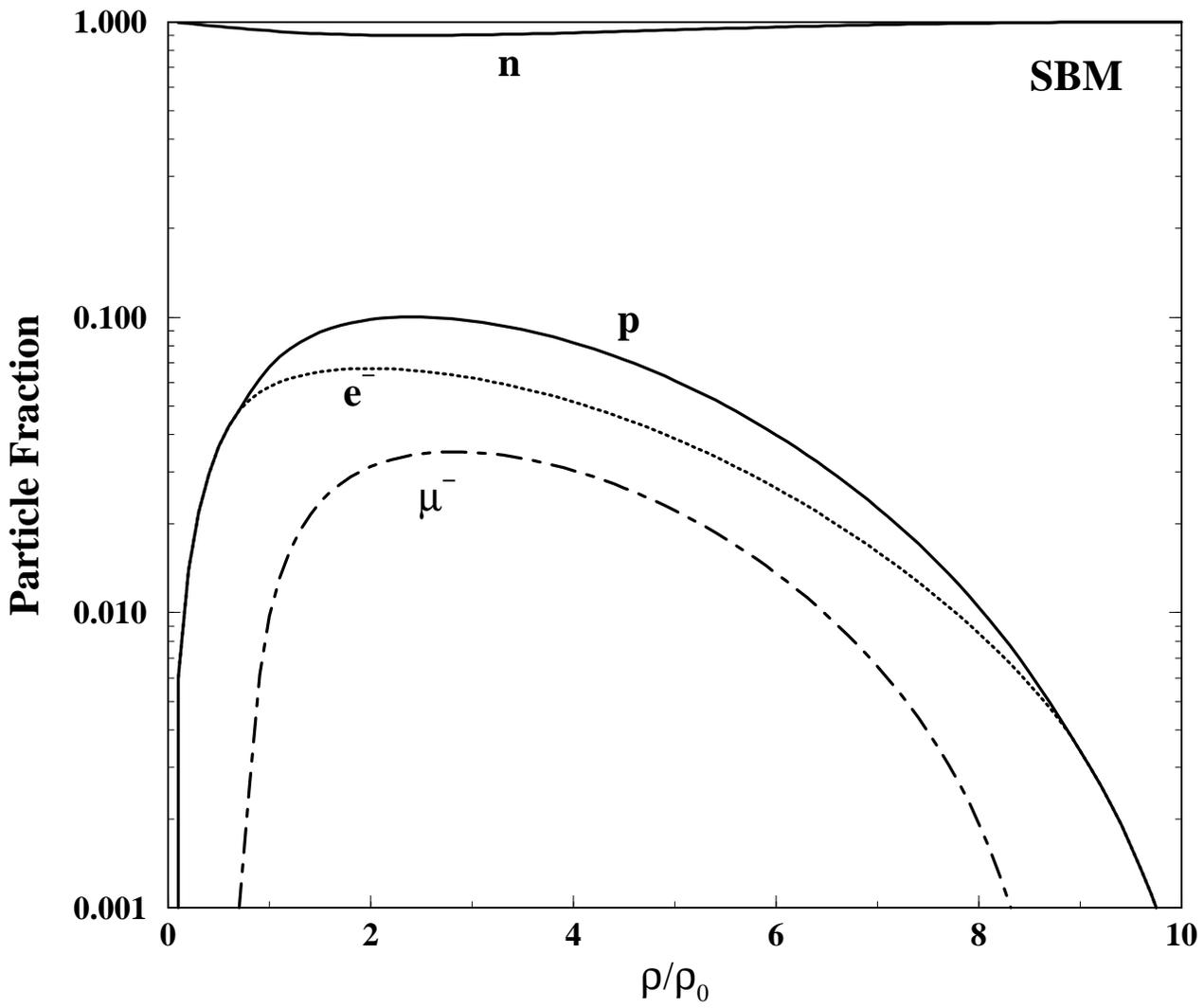
FIG. 1. Particle abundances of nucleons-only matter as a function of 
normalised baryon density.
\vspace{0.8cm}

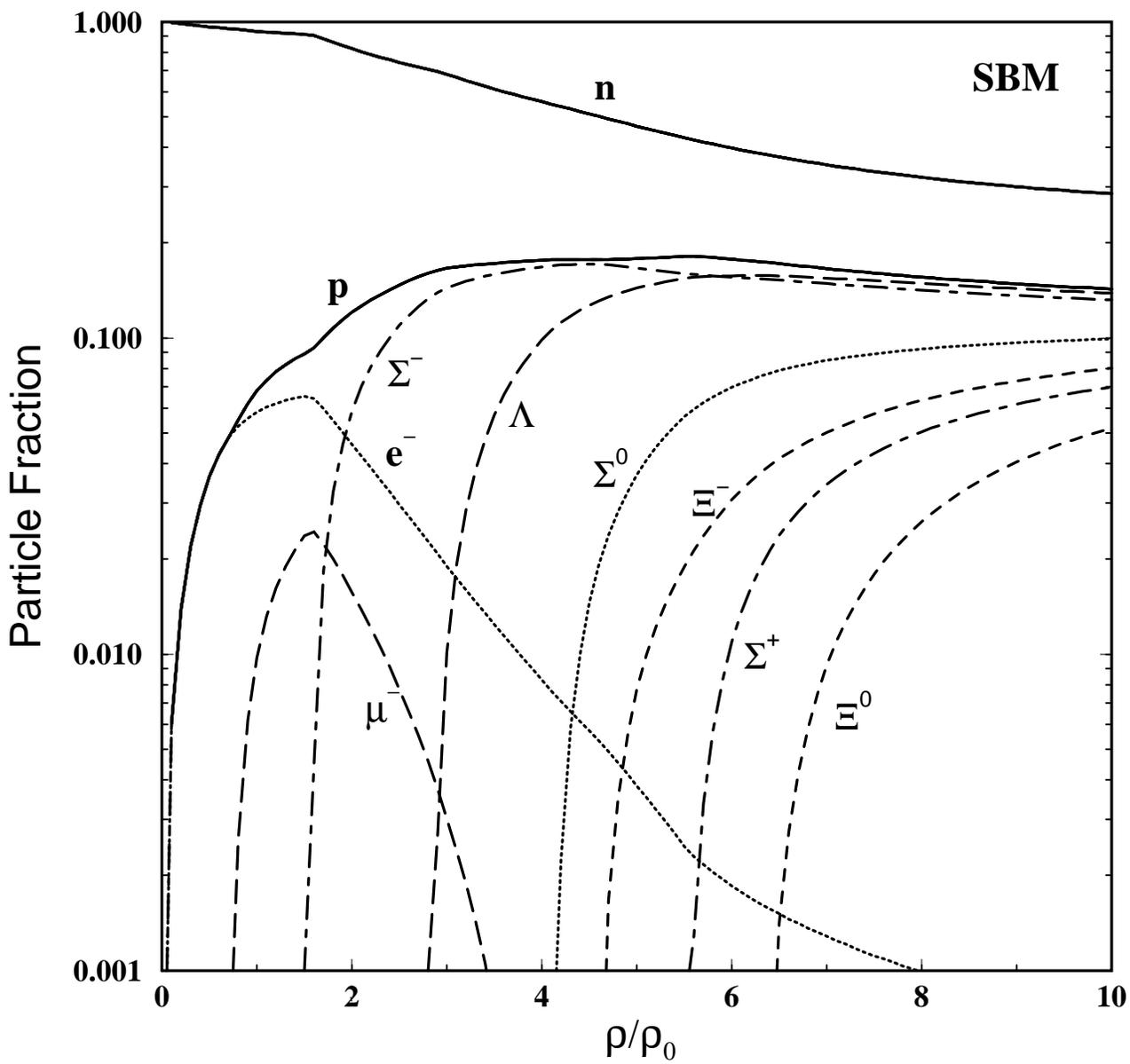
FIG. 2. Particle abundances of hyperon matter as a function of normalised 
baryon density.
\vspace{0.8cm}

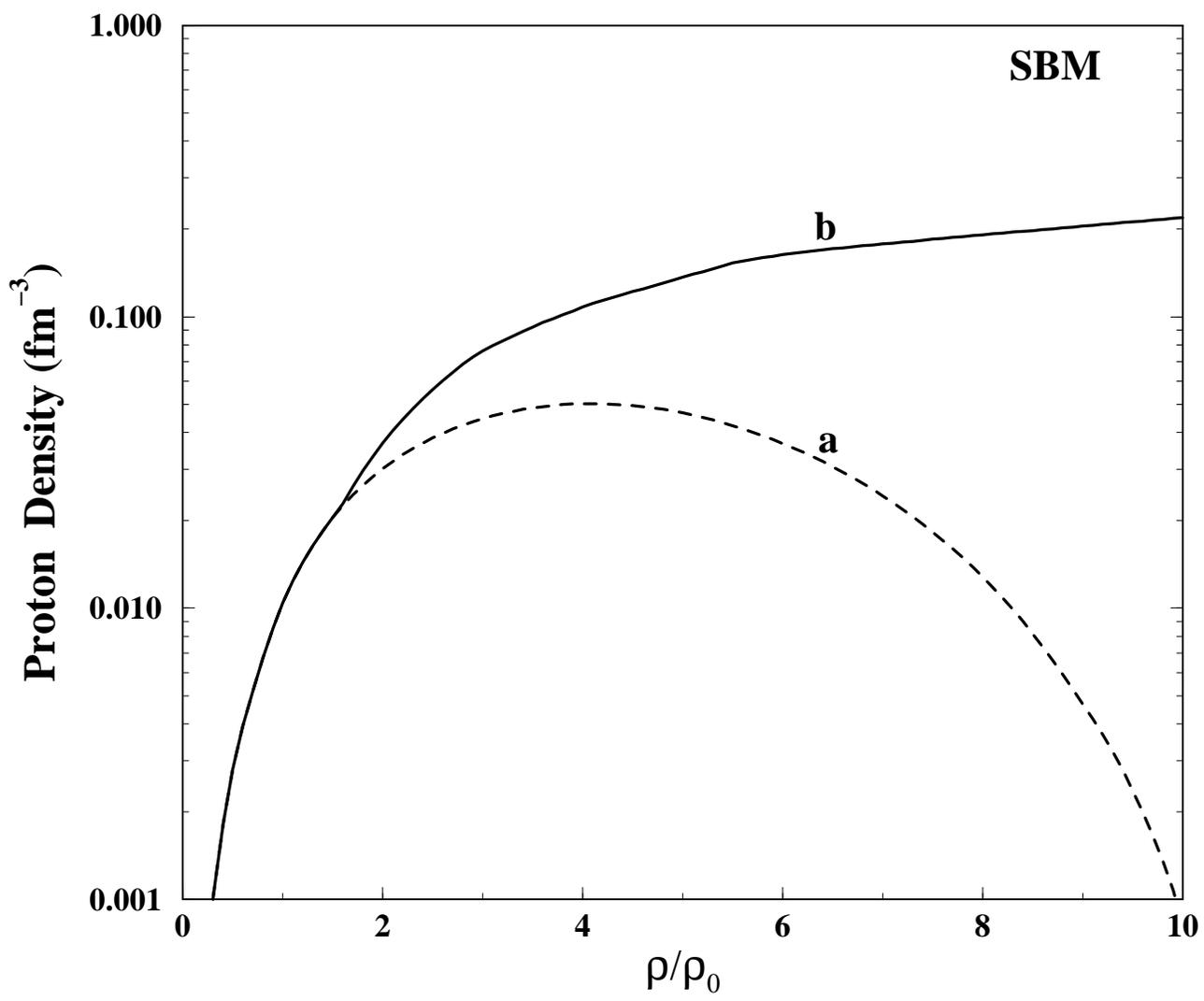
FIG. 3. Proton density as a function of normalised baryon density in  
nucleons-only and hyperon matter.
\vspace{0.8cm}

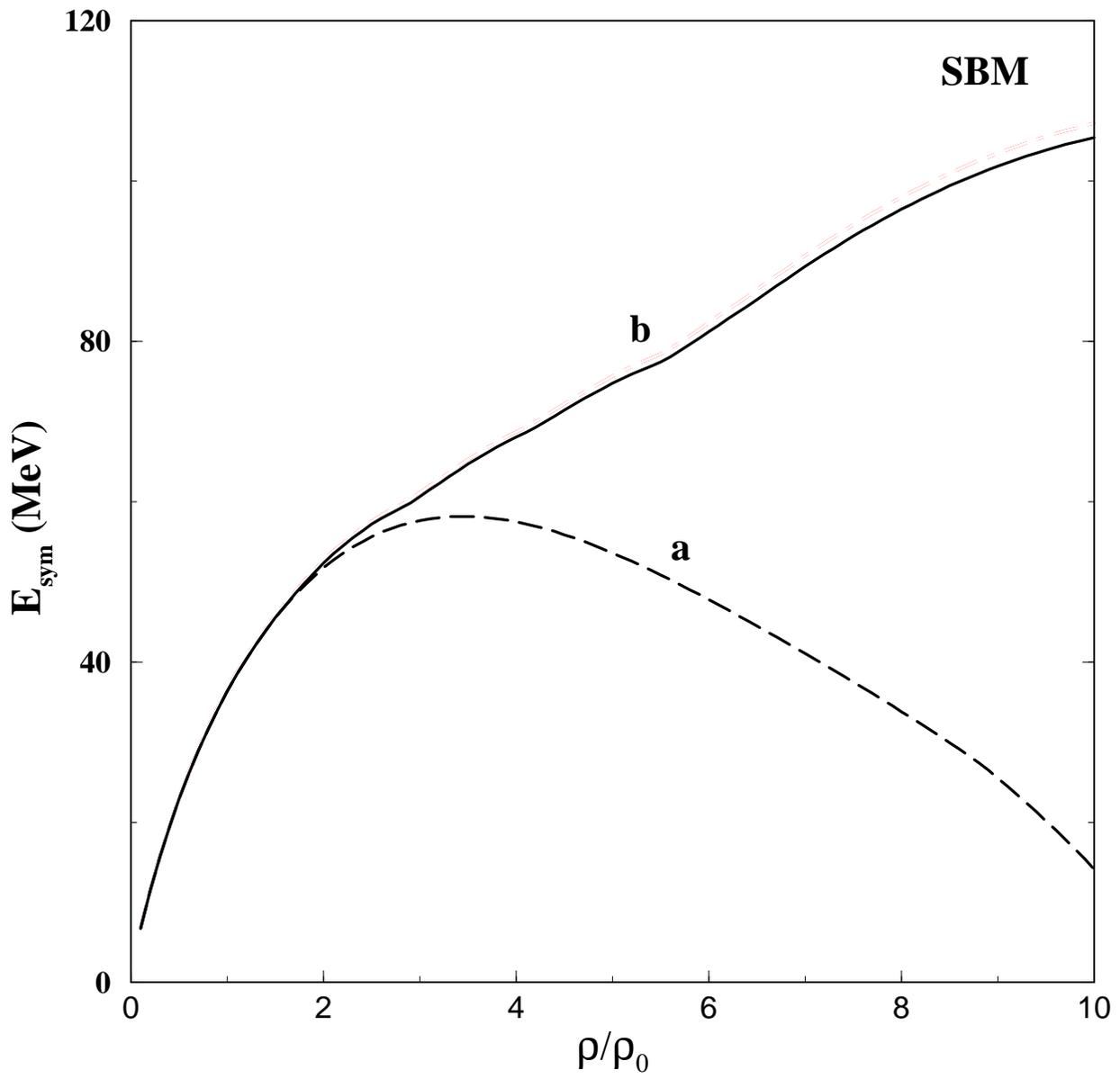
FIG. 4. Nuclear symmetry energy as a function of normalised baryon density in  
nucleons-only and hyperon matter.
\end{document}